\documentclass[a4paper,12pt]{article}

\usepackage{amsmath}
\usepackage[psamsfonts]{amssymb}
\usepackage{rsfs}
\usepackage{bm}

\usepackage{cite}
\usepackage[dvips]{graphicx}
\usepackage{color}

\makeatletter
\@addtoreset{equation}{section}
\makeatother

\begin{document} 

\begin{titlepage}

\baselineskip 10pt
\vskip 5pt
\leftline{}
\leftline{Chiba Univ. Preprint
          \hfill   \small \hbox{\bf CHIBA-EP-178}}
\leftline{\hfill   \small \hbox{September 2009}}
\vskip 5pt
\baselineskip 14pt
\centerline{\Large\bf 
} 
\vskip 0.2cm
\centerline{\Large\bf  
Decoupling and scaling solutions 
}
\vskip 0.2cm
\centerline{\Large\bf  
in Yang-Mills theory with the Gribov horizon
}
\vskip 0.2cm

\vskip 0.2cm

\centerline{{\bf 
Kei-Ichi Kondo,$^{\dagger,{1},{2}}$  
}}  
\vskip 0.2cm
\centerline{\it
${}^{1}$Department of Physics, University of Tokyo,
Tokyo 113-0033, Japan
}
\centerline{\it
${}^{2}$Department of Physics, 
Chiba University, Chiba 263-8522, Japan
}
\vskip 0.2cm

\begin{abstract}
We propose a trick which enables one to incorporate  the Gribov horizon into the Schwinger-Dyson equation  in Landau and Coulomb gauge Yang-Mills theory, using the Gribov-Zwanziger framework with the horizon term. 
We find a family of solutions parameterized by one-parameter $w_R(0)$ which was overlooked so far by assuming to be zero implicitly. 
The family includes both the scaling and decoupling solutions, and specification of the parameter  discriminates between them. In the Landau gauge we discuss a possible decoupling solution satisfying the Kugo-Ojima criterion for color confinement. 

\end{abstract}

 

\hrule  
${}^\dagger$ 
On sabbatical leave of absence from Chiba University. 
\\
 E-mail:  {\tt kondok@faculty.chiba-u.jp}

\par 
\par\noindent


\vskip 0.3cm

\pagenumbering{roman}
\tableofcontents




\end{titlepage}


\pagenumbering{arabic}

\baselineskip 14pt

\section{Introduction}

The Green functions are quantities of fundamental importance in quantum field theories.  In quantum Yang-Mills theory \cite{{YM54}}, the Green functions are   gauge-dependent quantities and defined only when the gauge fixing condition is imposed, as far as the local Green functions are concerned.  
In other words, the result of calculating the Green functions can be affected by details of the explicit procedure for gauge fixing, in sharp contrast to the gauge-invariant quantities which do not depend on the gauge fixing adopted and should take the same value even if they are calculated in different gauges. 
However, the gauge-invariant operators such as the Wilson loop operator and the Polyakov line operator are necessarily non-local.

The quantum Yang-Mills theory in the Landau gauge
$\partial \mathscr{A}=0$ as a manifestly covariant gauge is naively defined by the functional integral:
\begin{equation}
 Z_{\rm YM}  := \int [d\mathscr{A}] \delta(\partial \mathscr{A}) \det (-\partial D[\mathscr{A}]))\exp \{ -S_{YM}[\mathscr{A}]  \} 
 , 
\end{equation}
where $S_{YM}$ is the Yang-Mills action and $-\partial D[\mathscr{A}]$ is the Faddeev-Popov operator. 
However, the Landau gauge fixing $\partial \mathscr{A}=0$ can not fix the gauge uniquely. 
This is because each gauge orbit intersects the gauge fixing hypersurface $\Gamma := \{ \mathscr{A}; \partial \mathscr{A}=0 \}$ many times. The unique representative from each gauge orbit can not be chosen by imposing  $\partial \mathscr{A}=0$ alone. There are many representatives called the Gribov copies \cite{Gribov78}. 

In order to avoid the Gribov copies, Gribov \cite{Gribov78} proposed to restrict the functional integral to the 1st Gribov region $\Omega$: 
\begin{equation}
 Z_{\rm Gribov}  := \int_{\Omega} [d\mathscr{A}] \delta(\partial \mathscr{A}) \det (-\partial D[\mathscr{A}]))\exp \{ -S_{YM}[\mathscr{A}]  \} 
  ,
\end{equation}
where the Gribov region is defined by
\begin{equation}
 \Omega := \{ \mathscr{A}  ; \partial \mathscr{A} = 0 ,  \ -\partial D[\mathscr{A}] > 0 \} \subset \Gamma  
 .
\end{equation}
The 1st Gribov region is a bounded and convex region including the origin $\{ \mathscr{A}=0\}$. In fact,  $-\partial_\mu D_\mu[\mathscr{A}=0] = -\partial_\mu  \partial_\mu > 0$, i.e., $\{ \mathscr{A}=0\} \in \Omega$.
The boundary of $\Omega$ is called the Gribov horizon:
\begin{equation}
 \partial \Omega := \{ \mathscr{A} ; \partial \mathscr{A} = 0 , \ -\partial D[\mathscr{A}] = 0 \} 
  .
\end{equation}
He predicted that the resulting Green functions exhibit unexpected behavior in the deep infrared (IR) region and that they play the essential role in confinement.

We define the gluon 2-point function (full or complete propagator) by
\begin{equation}
 D_{\mu\nu}^{AB}(k) := \delta^{AB} \left[ \left(\delta_{\mu\nu}-\frac{k_\mu k_\nu}{k^2}  \right) \frac{F(k^2)}{k^2} + \frac{\alpha}{k^2} \frac{k_\mu k_\nu}{k^2} 
\right] \quad (\alpha=0) ,
\end{equation}
and the ghost propagator by
\begin{equation}
 G^{AB}(k) := - \delta^{AB} \frac{G(k^2)}{k^2}
 ,
\end{equation}
where the free case corresponds to  
$
F(k^2) \equiv 1
$
and
$
G(k^2) \equiv 1
$. 
 Gribov predicted their behaviors in the deep IR region $k^2 \ll 1$:
\begin{equation}
  \frac{F(k^2)}{k^2} \sim  \frac{k^2}{(k^2)^2+M^4} \downarrow 0 , \quad  G(k^2)  \sim \frac{M^2}{k^2} \uparrow \infty \quad (k^2  \downarrow 0) 
 , 
\end{equation}
where  $M$ is a constant with mass dimension called the Gribov mass. This power like behavior should be compared with the ultraviolet (UV)  behavior with the logarithmic corrections.

The Gribov prediction was investigated more elaborately  by solving the coupled Schwinger-Dyson (SD) equation for the gluon and ghost propagators. 
The dressing functions $F$ and $G$ are characterized by the power behavior with exponents $\alpha$ and $\beta$ respectively: 
\begin{align}
  F(k^2) = A \times (k^2)^{\alpha}, \quad G(k^2) = B \times (k^2)^{\beta} , \quad
  \alpha+2\beta=0, \quad 0 < A, B< \infty 
   .
\end{align}
By the scaling relation $\alpha+2\beta=0$, a single exponent $\kappa$ is enough for characterizing the IR behavior: 
\begin{equation}
\alpha = 2\kappa >1, \quad \beta= -\kappa <0, \quad
  \  1/2<\kappa<1 \quad (\text{Gribov} \ \kappa=1 ).
\end{equation}
The Gribov prediction corresponds to the limiting value $\kappa=1$.
This result, gluon suppression and ghost dominance, leads to the running coupling constant with a non-trivial IR fixed point:
\begin{align}
    g^2(k) :=  g^2 F(k^2)G^2(k^2)
  & \rightarrow (0 <) g^2 AB^2 (<\infty) 
 \quad (k^2 \rightarrow 0)  
  .
\end{align}
This solution is called the \textit{scaling solution}.
\textit{The gluon propagator $F(k^2)/k^2$ vanishes in the IR limit $k^2  \downarrow 0$, while the ghost propagator becomes more singular than the free case in the IR region, or the ghost dressing function $G(k^2)$ diverges}, i.e., 
\begin{equation}
 G(0)=\infty .
\end{equation} 
See the excellent review by Alkofer and von Smekal \cite{AS01} for details.

This IR behavior was considered to be reasonable from the viewpoint of color confinement.
Due to Kugo and Ojima \cite{KO79}, all color non-singlet objects can not be observed or confined, in other words, only color singlet objects are observed, if a criterion  $u(0)=-1$ is satisfied in the Lorentz covariant gauge 
(a sufficient condition for color confinement). 
\footnote{
Note that the Kugo-Ojima theory for color confinement   does not take into account the Gribov problem and is based on the usual BRST formulation where the exact color symmetry and the well-defined BRST charge are assumed.  
}
It is shown in \cite{Kugo95} that \textit{in the Landau gauge, the Kugo-Ojima criterion for color confinement $u(0)=-1$ is equivalent to the  divergent ghost dressing function $G(0)=\infty$ }, since  in the Landau gauge 
\begin{equation}
G(0)=[1+u(0)]^{-1}
 .
\end{equation}
In this paper, we point out that this relation is not exact and must be used with more care, since  
$u(0)=-1$ does not necessarily mean $G(0)=\infty$.

Until 2006, it seemed that the scaling solution has been confirmed by the SD equation, the functional renormalization group equation and numerical simulations on lattice.  This lead to the \textit{ghost dominance  picture for color confinement}.



So far so good. 
However,  these results are questioned by the Orsay group at Universite de Paris Sud.
By careful analyses of the SD equation,  socalled \textit{the decoupling solution} was discovered \cite{Boucaudetal08}: 
\begin{align}
 & F(k^2) = A^\prime \times (k^2)^{ \alpha^\prime }, \quad G(k^2) = B^\prime \times (k^2)^{ \beta^\prime } , \quad
   \alpha^\prime=1, \quad  \beta^\prime=0 , \quad 0 < A^\prime, B^\prime< \infty  ,
\end{align}
which lead to the running coupling going to zero in the IR limit:
\begin{equation}
    g^2(k) :=  g^2 F(k^2)G^2(k^2)
    \sim  g^2 A^\prime B^\prime{}^2 k^2  \rightarrow 0
 \quad (k^2 \rightarrow 0) , 
\end{equation}
although its possibility was mentioned also in \cite{LS02}.
Moreover, reexaminations of numerical simulations on large lattices \cite{SIMPS06,BMMP08,BIMPS09,CM07,CM08,OS08,SS08}, functional renormalization group equation \cite{FMP08} and other methods \cite{ABP08,Dudaletal08} seem to support this result.
\textit{The decoupling solution implies that the gluon propagator goes to the non-zero and finite constant in the IR limit, while  the ghost propagator behaves like free, namely, the ghost dressing function G(0) is non-zero and finite in the IR limit}:
\begin{equation}
 0<G(0)<\infty . 
\end{equation}
In the decoupling solution, the gluon decouples below its mass scale and the ghost is still dominant, although the ghost dominance in the decoupling solution is weaker than that in the scaling solution.

In this paper, we take into account the existence of the Gribov horizon by making use of the Gribov-Zwanziger theory \cite{Zwanziger89,Zwanziger92,Zwanziger93}:   
\begin{equation}
 Z_{\rm GZ} := \int [d\mathscr{A}] \delta (\partial^\mu \mathscr{A}_\mu) \det (K[\mathscr{A}]) \exp \{ -S_{YM}[\mathscr{A}] - \gamma \int d^D x h(x) \} 
 , 
 \label{YM1}
\end{equation}
where $K$ is the Faddeev-Popov operator $K[\mathscr{A}]:=-\partial_\mu D_\mu=-\partial_\mu (\partial_\mu+g \mathscr{A}_\mu \times)$ and $h(x)=h[\mathscr{A}](x)$ is the Zwanziger \textit{horizon function}.
Here the parameter $\gamma$ called the \textit{Gribov parameter} is determined by solving a gap equation, commonly called the \textit{horizon condition}: for $D$-dimensional Euclidean SU(N) Yang-Mills theory,
\begin{equation}
 \langle h(x) \rangle_{\rm GZ} = (N^2-1)D .
\end{equation}
The horizon function plays the role of restricting the integration region inside the Gribov horizon.  However, the exact form of the horizon function in this sense is not known and the choice of the horizon function is not unique at present. Some arguments on this point are given later. 
The first choice of the horizon function is \cite{Zwanziger89}  
\begin{equation}
 h(x) 
=  \int d^Dy gf^{ABC} \mathscr{A}_\mu^{B}(x) (K^{-1})^{CE}(x,y) gf^{AFE} \mathscr{A}_\mu^{F}(y)
 . 
 \label{h1}
\end{equation}
The second choice is \cite{Zwanziger93}
\begin{equation}
  h(x) =  \int d^Dy D_\mu[\mathscr{A}]^{AC}(x) (K^{-1})^{CE}(x,y) D_\mu[\mathscr{A}]^{AE}(y) .
  \label{h2}
\end{equation}
In any case, inclusion of the  horizon term makes the theory non-local.

The SD equations for Green functions do not change their form even in the presence of the Gribov horizon, since the integrand of the functional integration formula for Green functions vanishes at the Gribov horizon which is the boundary of the functional integration region due to the Faddeev-Popov operator. 
Hence, the solutions of the SD equation include both the solution with the Gribov horizon and the solution without restriction.   
Remarkably, it has been shown  \cite{FMP08,FMP07} that the set of solutions of  the coupled SD equation for gluon and ghost propagators is uniquely determined once a boundary value $G(0)$ is given, corresponding to the scaling solution for $G(0)=\infty$ and the decoupling solution for $0<G(0)<\infty$. 
However, it is not yet examined how these solutions are related to the Gribov horizon. 
Moreover, the horizon term does not uniquely fix the gauge, since there are still Gribov copies in the Gribov region. 
A one-parameter family of correlation functions are constructed in lattice gauge theory  distinguished by a second gauge parameter $B$ (Landau-$B$ gauge) \cite{Maas09}. This uniquely specifies a representative from a gauge orbit and no further freedom in choosing a Gribov copy. 
In this paper, we give a trick to incorporate the horizon condition into the SD equation of the ghost propagator, which enables us to distinguish the solution associated with the particular choice of the horizon term. 
This is possibly used to discrimate the scaling and decoupling solutions. 
 We show that both horizon terms allow the existence of one-parameter family of solutions parameterized by a real number $w_R(0)$ which has been assumed implicitly to be zero $w_R(0)=0$ in the previous investigations.  Therefore, $w_R(0)$ plays the role of an additional non-perturbative gauge parameter which uniquely specifies the solution. 
 We consider both the unrenormalized and renormalized versions of the SD equation with the horizon condition being included. 
 
\section{Schwinger-Dyson equation with the horizon condition being inserted}

In what follows, we consider the SU(N) Yang-Mills theory in $D$-dimensional Euclidean space. 

The Schwinger-Dyson (SD) equation for the ghost propagator $\langle   \mathscr{C}^A \bar{\mathscr{C}}^B \rangle_k$ in   momentum space is written in the following form. 
We follow the notation of \cite{Kondo09a,Kondo09c}.
\begin{equation}
   \langle   \mathscr{C}^A \bar{\mathscr{C}}^B \rangle_k^{-1} = - \delta^{AB} k^2 
-i \frac{k^\mu}{k^2} \langle (g \mathscr{A}_\mu \times \mathscr{C})^A \bar{   \mathscr{C}}^B \rangle_k^{\rm 1PI} 
 ,
\end{equation}
which  is obtained as the Fourier transform of  
\begin{equation}
  0 =  - \langle  (\partial_\mu D_\mu[\mathscr{A}] \mathscr{C})^A(x) \bar{\mathscr{C}}^B(y) \rangle   + \delta^{AB}  \delta^D(x-y) 
 .
\end{equation} 
In the Gribov theory, this is derived from the identity:
\begin{equation}
 0 = \int_{\Omega} [d\mathscr{A}] [d\mathscr{B}][d\mathscr{C}][d\bar {\mathscr{C}}] 
\frac{\delta}{\delta \bar{\mathscr{C}}^A(x)} \left[  e^{ -S_{\rm YM}^{\rm tot}  } \bar{\mathscr{C}}^B(y) \right]
 , 
\end{equation}
while in the Gribov-Zwanziger theory, the same form is obtained   from 
\begin{equation}
 0 = \int [d\mathscr{A}] [d\mathscr{B}][d\mathscr{C}][d\bar {\mathscr{C}}] 
\frac{\delta}{\delta \bar{\mathscr{C}}^A(x)} \left[  e^{ -S_{\rm YM}^{\rm tot} - \gamma \int d^D x h(x) } \bar{\mathscr{C}}^B(y) \right]
 , 
\end{equation}
where
\begin{align}
 S_{\rm YM}^{\rm tot}  :=& S_{\rm YM} + S_{\rm GF+FP} ,
 \nonumber\\
 S_{\rm YM} :=&  \int d^Dx \frac14 \mathscr{F}_{\mu\nu}  \cdot \mathscr{F}_{\mu\nu} ,
 \nonumber\\
 S_{\rm GF+FP} :=& \int d^Dx  \left\{ \mathscr{B} \cdot \partial_\mu \mathscr{A}_\mu 
+i \bar {\mathscr{C}} \cdot \partial_\mu D_\mu \mathscr{C} \right\} 
 ,
\end{align}
and the dot and the cross are defined as 
$
 \mathscr{A} \cdot \mathscr{B} := \mathscr{A}^A \mathscr{B}^A
$
and
$
 (\mathscr{A} \times \mathscr{B})^A := f^{ABC} \mathscr{A}^B \mathscr{B}^C 
$.

\begin{figure}
\begin{center}
\includegraphics[width=6.0in]{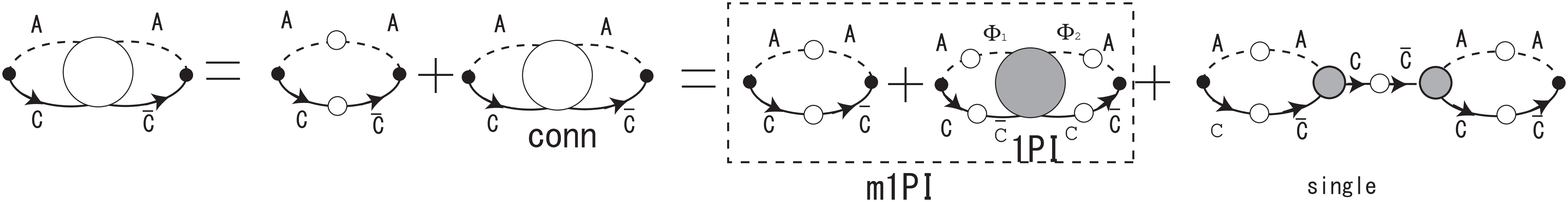}
\end{center}
\caption{Diagrammatic representation of 
$\langle   (g \mathscr{A}_\mu \times \mathscr{C})^A (g \mathscr{A}_\nu \times  \bar{\mathscr{C}})^B \rangle_k$,
$\langle   (g \mathscr{A}_\mu \times \mathscr{C})^A (g \mathscr{A}_\nu \times  \bar{\mathscr{C}})^B \rangle_k^{\rm conn}$, 
$\langle   (g \mathscr{A}_\mu \times \mathscr{C})^A (g \mathscr{A}_\nu \times  \bar{\mathscr{C}})^B \rangle_k^{\rm 1PI}$
and 
$\langle   (g \mathscr{A}_\mu \times \mathscr{C})^A (g \mathscr{A}_\nu \times  \bar{\mathscr{C}})^B \rangle_k^{\rm m1PI}$.
}
\label{fig:4p-def-diagram}
\end{figure}

By using the relationship derived in \cite{Kondo09a,Kondo09c}
\begin{equation}
 -i \frac{k^\mu}{k^2} \langle (g \mathscr{A}_\mu \times \mathscr{C})^A \bar{   \mathscr{C}}^B \rangle_k^{\rm 1PI}
=   \frac{k^\mu k^\nu}{k^2} \lambda_{\mu\nu}^{AB}(k)  
   = \delta^{AB}[u(k^2)+w(k^2) ]
  ,
\end{equation} 
the SD equation for the ghost dressing function $G(k^2)$ is rewritten as \cite{Kondo09a,Kondo09c}
\begin{equation}
  \quad 
  G^{-1}(k^2) =  {1} + u(k^2)+w(k^2) 
  .
\end{equation}
This identity was derived by \cite{Kugo95} and also in \cite{GHQ04} based on a different method.
Here two functions $u$ and $w$ are defined from the modified 1-particle irreducible (m1PI) part as
\begin{equation}
 \lambda_{\mu\nu}^{AB}(k) := \langle   (g \mathscr{A}_\mu \times \mathscr{C})^A (g \mathscr{A}_\nu \times  \bar{\mathscr{C}})^B \rangle_k^{m1PI}
=  \left[ \delta_{\mu\nu} u(k^2) + \frac{k_\mu k_\nu}{k^2} w(k^2) \right] \delta^{AB} 
 ,
\end{equation}
where $u(k^2)$ agrees with the Kugo-Ojima function usually defined by 
\begin{equation}
  \langle  (D_\mu \mathscr{C})^{A}  (g\mathscr{A}_\nu \times \bar{\mathscr{C}})^{B}  \rangle_{k} 
:= \left(  \delta_{\mu\nu} - \frac{k_\mu k_\nu}{k^2} \right) \delta^{AB} u(k^2)   
 . 
\end{equation}
The m1PI part is defined from the two-point function of the composite operators (See Fig.~\ref{fig:4p-def-diagram})
\begin{equation}
   \langle   (g \mathscr{A}_\mu \times \mathscr{C})^A (g \mathscr{A}_\nu \times  \bar{\mathscr{C}})^B \rangle_k
=    \lambda_{\mu\nu}^{AB}(k)
+  \Delta_{\mu\nu}^{AB}(k)
 ,
\end{equation}
where
\begin{align}
\lambda_{\mu\nu}^{AB}(k) :=&  \langle   (g \mathscr{A}_\mu \times \mathscr{C})^A (g \mathscr{A}_\nu \times  \bar{\mathscr{C}})^B \rangle_k^{\rm m1PI}
,
\nonumber\\
 \Delta_{\mu\nu}^{AB}(k)   :=&  \langle (g \mathscr{A}_\mu \times \mathscr{C})^A \bar{   \mathscr{C}}^C \rangle_k^{\rm 1PI} 
 \langle \mathscr{C}^C \bar{\mathscr{C}}^D \rangle_k 
 \langle \mathscr{C}^D (g \mathscr{A}_\nu \times \bar{   \mathscr{C}})^B \rangle_k^{\rm 1PI}   
   .
\end{align}


Following the idea of Gribov \cite{Gribov78}, we incorporate the horizon condition 
$
 \langle  h(0) \rangle = (N^2-1)D
$
into the SD equation. 
By substituting the horizon condition into the free (tree) part of the SD equation, the SD equation for the ghost dressing function reads
\begin{equation}
  G^{-1}(k^2) =   \frac{\langle h(0) \rangle}{(N^2-1)D}  + u(k^2)+w(k^2) 
  ,
  \label{SDwHa}
\end{equation}
which is equivalent to the SD equation for the ghost propagator:
\begin{equation}
   \langle   \mathscr{C}^A \bar{\mathscr{C}}^B \rangle_k^{-1} = - \delta^{AB} k^2 
 \frac{\langle h(0) \rangle}{(N^2-1)D} 
-i \frac{k^\mu}{k^2} \langle (g \mathscr{A}_\mu \times \mathscr{C})^A \bar{   \mathscr{C}}^B \rangle_k^{\rm 1PI} 
 .
  \label{SDwHb}
\end{equation}

\subsection{The first horizon term}

The first horizon term (\ref{h1}) yields \cite{Kondo09a,Kondo09c}  
\begin{align}
 \langle  h(0) \rangle
  =& - \lim_{k \to 0}  \langle   (g \mathscr{A}_\mu \times \mathscr{C})^A (g \mathscr{A}_\mu \times  \bar{\mathscr{C}})^A \rangle_k
  \nonumber\\
=& 
 - (N^2-1)  \left\{ Du(0) +w(0)- G(0) [u(0)+w(0)]^2  \right\}  
 ,
 \label{hc1}
\end{align} 
and
\begin{equation}
  \frac{\langle  h(0) \rangle}{(N^2-1)D}
= -u(0)-\frac{w(0)}{D}+ \frac{G^{-1}(0)-2+G(0)}{D}
 .
\end{equation}
Then the SD equation reads 
\begin{equation}
  G^{-1}(k^2) =     \frac{G^{-1}(0)-2+G(0)}{D} -\frac{w(0)}{D} -u(0) + u(k^2)+w(k^2) 
  .
\end{equation}
Then, it is observed that the  term $-u(0)$ coming from the horizon condition exactly cancels the self-energy term $u(k^2)$ at $k=0$ in the right-hand side of the SD equation.  
In the deep IR limit $k=0$,  therefore, we have
\begin{equation}
  G^{-1}(0) =  
 \frac{G^{-1}(0)-2+G(0)}{D}
+ \left(  - \frac{1}{D}  + 1  \right) w(0) 
  .
  \label{ghostSDEda2}
\end{equation}
By solving this equation for $G(0)$:
\begin{equation}
  G^{2}(0) -[2-(D-1)w(0)]G(0) + 1-D = 0 ,
\end{equation}
we have
\begin{equation}
G(0) = 1 -(D-1)w(0)/2+ \sqrt{[1-(D-1)w(0)/2]^2-1+D} > 0
 ,
\end{equation} 
and $u(0) =  -1-w(0)+G^{-1}(0)$, i.e., 
\begin{equation}
u(0) 
= -1-w(0)- \frac16 \{ 2-3w(0)-\sqrt{12+[2-3w(0)]^2} \}
 .
\end{equation} 
This implies  that \textit{the horizon condition determines the boundary value $G(0)$ in the ghost SD equation. 
Consequently, we have one-parameter family of solutions parameterized by $w(0)$.}



We consider $G(0)$ and $u(0)$ as  functions of $w(0)$.
Both $G(0)$ and $u(0)$ are monotonically decreasing functions in $w(0)$;    $G(0), u(0) \rightarrow \infty$ as $w(0) \rightarrow -\infty$, while $G(0) \rightarrow 0$ and $u(0) \rightarrow -5/3$ as $w(0)  \rightarrow +\infty$.  For $D=4$, in particular, $G(0)=3$ and $u(0)=-2/3$ at $w(0)=0$; $G(0)=2$ and $u(0)=-1$ at $w(0)=1/2$.

Thus the scaling solution $G(0)=\infty$ is obtained only when $w(0)=-\infty$. 
Otherwise $w(0)>-\infty$, the decoupling solution $0< G(0) < \infty$ is obtained. 
Using a special value as an additional input $w(0)=0$ by an assumption \cite{Kugo95,Kondo09a} or  by an independent argument \cite{ABP09}
 $G(0)$ is determined selfconsistently by solving the above SD equation as \cite{Kondo09a,Kondo09c}
\begin{equation}
G(0) = 1 + \sqrt{D} > 0,
\quad
u(0)=(-D \pm \sqrt{D})/(D-1)
 .
\end{equation} 
In particular, for $D=4$, 
\begin{equation}
G(0) = 3 > 0,
\quad
u(0)=-2/3 \quad (D=4)
 .
\end{equation}

In order to obtain the scaling solution, the constant terms must cancel exactly or disappear at the $k=0$ limit on the right-hand side of the SD equation. 
This is what implicitly assumed, but not stated explicitly, 
as pointed out by \cite{Boucaudetal08}.

\subsection{The second horizon term}

The second horizon function (\ref{h2}) leads to
\begin{align}
 \langle  h(0) \rangle
=  -\lim_{k  \rightarrow 0} \langle   (D_\mu  \mathscr{C})^A (D_\mu   \bar{\mathscr{C}})^A \rangle_{k} 
= -(N^2-1) 
  \left\{ (D-1)u(0) -1 \right\}  
 ,
 \label{hc2}
\end{align} 
and
\begin{align}
 \frac{\langle  h(0) \rangle}{(N^2-1)D}
=  -\frac{\lim_{k  \rightarrow 0} \langle   (D_\mu  \mathscr{C})^A (D_\mu   \bar{\mathscr{C}})^A \rangle_{k}}{(N^2-1)D} 
=    \frac{1}{D} [1+   u(0)] - u(0) 
 .
\end{align} 
Then the  SD equation  
$
  G^{-1}(k^2) =  1  + u(k^2)+w(k^2) 
$
is rewritten as
\begin{equation}
  G^{-1}(k^2) =   \frac{1}{D} [1+   u(0)] - u(0)  + u(k^2)+w(k^2) 
  .
\end{equation}
In the deep IR limit, the  cancellation occurs for   $u(0)$ in the right-hand side of this equation:
\begin{equation}
  G^{-1}(0) =   \frac{1}{D} [1+   u(0)]   +w(0) 
  ,
\end{equation}
which is rewritten   in favor of $G(0)$ using $G(0)^{-1}=1+u(0)+w(0)$  as
\begin{equation}
  G^{-1}(0) = \frac{1}{D} G^{-1}(0) -   \frac{1}{D} w(0)     +w(0) 
  .
\end{equation}
This is solved for $D \ne 1$ to give 
\begin{equation}
  G(0) =  w^{-1}(0) 
  ,
\end{equation}
and
\begin{equation}
  u(0) =  -1 
  .
  \label{KOc}
\end{equation}
Thus, we have one-parameter family of solutions parameterized by $w(0)$. 
The scaling solution $G(0)=\infty$ is obtained only when $w(0)=0$. 
Otherwise $0 < w(0) < \infty$, the decoupling solution $\infty >  G(0) > 0$ is obtained. 
It should be remarked that the Kugo-Ojima condition $u(0)=-1$ is always satisfied.
\footnote{
The fact that the horizon condition using this definition (\ref{hc2}) is equivalent to the Kugo-Ojima criterion (\ref{KOc}) has already been pointed out and it was checked to what extent the horizon condition holds in the numerical simulation in \cite{NF00}.  
}
 However, this does not immediately mean the enhancement of the ghost propagator, contrary to the usual claim found in  literatures.

If we require that two horizon conditions (\ref{hc1}) and (\ref{hc2}) give the same result, then the relation
\begin{equation}
 u(0)=-1/2-w(0)
\end{equation}
must be satisfied for any $D$.
This implies that the common solution is found for any $D$
\begin{equation}
  G(0)  =2, \quad   u(0) =  -1 , \quad w(0)= \frac12  
  .
\end{equation}

\section{Removing ultraviolet divergence and renormalization}

We have calculated the first horizon condition using 
\begin{equation}
 \langle  h(x) \rangle 
 = - \lim_{k \to 0}  \langle   (g \mathscr{A}_\mu \times \mathscr{C})^A (g \mathscr{A}_\mu \times  \bar{\mathscr{C}})^A \rangle_k
  .
\end{equation}
However, the composite operator $(g \mathscr{A}_\mu \times \mathscr{C})$ is not multiplicative renormalizable due to operator mixing:
\begin{equation}
 g \mathscr{A}_\mu \times \mathscr{C} =  Z_{C}^{-1/2}  (g \mathscr{A}_\mu \times \mathscr{C})_R +  Z_{C}^{-1/2}(1-Z_{C}) \partial_\mu \mathscr{C}_R 
 ,
 \label{AC-ren}
\end{equation}
where $\mathscr{C} =Z_{C}^{1/2} \mathscr{C}_R$.
This implies that in the first choice one can not use the conventional framework of multiplicative renormalization for removing the ultraviolet divergence. 
Nevertheless, the SD equation for the ghost propagator is form-invariant under the multiplicative renormalization:
\begin{equation}
   \langle   \mathscr{C}^A_R \bar{\mathscr{C}}_R^{B}  \rangle_{k} = - \delta^{AB}\frac{1}{k^2}
-i \frac{k^\mu}{k^2} \langle (g \mathscr{A}_\mu \times \mathscr{C})^A_R \bar{\mathscr{C}}^{B}_R  \rangle_{k}  
 .
\label{SD1-ren}
\end{equation} 
We will return to the renormalization problem of the first horizon term later after discussing the second horizon term.

\subsection{The second horizon term}

It is well known that the composite operator$D_\mu[\mathscr{A}] \mathscr{C}$ is multiplicative renormalizable
\begin{equation}
 D_\mu[\mathscr{A}] \mathscr{C} = Z_{C}^{-1/2} (D_\mu[\mathscr{A}] \mathscr{C})_R
  .
 \label{DC-ren} 
\end{equation} 
Hence the second horizon condition can be tamed by the conventional method:
\begin{align}
 \langle  h(x) \rangle
=  -\lim_{k  \rightarrow 0} \langle   (D_\mu[\mathscr{A}]  \mathscr{C})^A (D_\mu[\mathscr{A}]   \bar{\mathscr{C}})^A \rangle_{k} 
 .
\end{align}

The unrenormalized horizon condition  
$
 0 =  (N^2-1)D -  \langle  h(0) \rangle
$ 
is
\begin{align}
0 = (N^2-1)D -  \langle  h(0) \rangle  
=   (N^2-1)  (D-1)[1+ u(0) ] 
 .
\end{align} 
We can obtain the  multiplicatively renormalized horizon condition as
\begin{align}
0 =&   Z_{C}  [(N^2-1)D - \langle  h(0) \rangle ]
=   (N^2-1)   (D-1) Z_{C}[1+ u(0) ] 
\nonumber\\
=&  (N^2-1)   (D-1) [1+ u_R(0) ] 
 ,
\end{align} 
by adopting the renormalization relation:
\begin{align}
  1+ u(0)  
= Z_{C}^{-1} [1+ u_R(0) ] 
 .
\end{align} 
Therefore, we have the renormalized SD equation:
\begin{equation}
  G^{-1}_R(0) =  \frac{1}{D} [1+u_R(0)]     +w_R(0) 
  ,
\end{equation}
which is rewritten   in favor of $G(0)$ as
\begin{equation}
  G^{-1}_R(0) =   \frac{1}{D} G^{-1}_R(0) -   \frac{1}{D} w_R(0)   +w_R(0) 
  ,
\end{equation}
where we have used the renormalization:
\begin{equation}
  G^{-1}(0) = Z_{C}^{-1} G^{-1}_R(0), \quad 
 w(0) = Z_{C}^{-1}w_R(0) 
  ,
\end{equation}
This renormalization prescription is compatible with the renormalized relation:
\begin{equation}
  G^{-1}_R(k^2) =  {1} + u_R(k^2)+w_R(k^2) 
  .
\end{equation}
Thus the renormalized horizon condition is satisfied  only when
\begin{equation}
 u_R(0)=-1, \quad G_R^{-1}(0)=w_R(0) .
\end{equation}  
This is also obtained as a self-consistent solution of the SD equation. 

The scaling solution $G_R(0)=\infty$ follows immediately from  $w_R(0) = 0$ which was  assumed implicitly by previous studies.  
On the other hand, the decoupling solution $G_R(0)<\infty$ as a result of $w_R(0) \ne 0$ means the existence of the massless pole in the correlation function $\lambda_{\mu\nu}(k)$. 
In fact, $w_R(0) = 0$ is true at the tree level and holds in perturbation theory, since the massless pole coming from the elementary Faddeev-Popov ghost was already removed in the definition of the 1PI function $\lambda_{\mu\nu}(k)$. 
For $w_R(0) \ne 0$ to be realized, therefore, a massless boundstate must be formed anew in the channel by the non-perturbative effect and the massless pole must be generated. 
However, the existence of such a pole would not mean the appearance of the physical massless particle in the spectrum, since it is not a gauge-invariant object. 
Therefore, $w_R(0) \ne 0$ does not contradict with the experiment. 
This consideration leads to a possibility of the decoupling solution satisfying the Kugo-Ojima criterion for color confinement.

\subsection{The first horizon term}

We return to the first horizon term
\begin{equation}
\langle  h(x) \rangle 
 = - \lim_{k \to 0}  \langle   (g \mathscr{A}_\mu \times \mathscr{C})^A (g \mathscr{A}_\mu \times  \bar{\mathscr{C}})^A \rangle_k
  ,
\end{equation}
which leads to the unrenormalized horizon condition
\begin{align}
0 =& (N^2-1)D -  \langle  h(0) \rangle 
\nonumber\\
=&    (N^2-1)  \left\{ D[1+u(0)] +w(0) - G(0)[u(0)+w(0)]^2 \right\} 
 .
\end{align} 
However, this horizon condition is not multiplicatively renormalizable!
\begin{align}
0 =&  Z_{C}[(N^2-1)D - \langle  h(0) \rangle ]
\nonumber\\
=&  (N^2-1)DZ_{C} + \lim_{k  \rightarrow 0}  Z_{C} \langle   (g \mathscr{A}_\mu \times \mathscr{C})^A (g \mathscr{A}_\mu \times  \bar{\mathscr{C}})^A \rangle_k   
\nonumber\\
=&    (N^2-1)  \left\{ DZ_{C}[1+u(0)] +Z_{C} w(0) - Z_{C} G(0)[u(0)+w(0)]^2 \right\} 
\nonumber\\
\ne&    (N^2-1)  \left\{ D [1+u_R(0)] +  w_R(0) -   G_R(0)[u_R(0)+w_R(0)]^2 \right\}  
 .
\end{align} 
Note that the renormalization prescription 
\begin{align}
  1+ u(0)  
=  Z_{C}^{-1} [1+ u_R(0) ] , 
\quad  w(0) = Z_{C}^{-1}w_R(0) , 
\quad
  G^{-1}_R(0) =& Z_{C}^{-1} G^{-1}(0) 
 ,
\end{align}
can not maintain the first horizon condition. 

\newpage
\underline{The conventional method:}
This is reasonable, since we did not use the localized renormalizable GZ theory.  Without introducing the Zwanziger ghost field $\xi, \bar\xi, \omega, \bar\omega$, the usual framework of the multiplicative renormalization does not work. 
Therefore, we must move to the  localized  GZ theory which is  manifestly multiplicative renormalizable.  
The Slavnov-Taylor identity means the horizon condition is translated to  (See Appendix A)
\begin{equation}
 \langle i\gamma^{-1/2}  g^2 f^{ABC}\mathscr{A}_\mu^B(x) \bar{\xi}^{CA}_\mu(x)   \rangle
 = \langle   h(x)  \rangle =  (N^2-1)D  
  .
\end{equation} 
This is rewritten into the covariant derivative form, since the average does not depend on $x$ due to the translational invariance: 
\begin{equation}
 \langle i\gamma^{-1/2}  g D_\mu[\mathscr{A}]^{AC} \bar{\xi}^{CA}_\mu    \rangle
 = \langle   h(0)  \rangle =  (N^2-1)D  
  .
\end{equation} 
Then the horizon condition is multiplicatively  renormalized: 
\begin{equation}
 \langle i\gamma_{R}^{-1/2}  g_{R} (D_\mu[\mathscr{A}]^{AC} \bar{\xi}^{CA}_\mu )_{R}  \rangle
 =Z_{C}   \langle   h(0)  \rangle = Z_{C}  (N^2-1)D  
  .
\end{equation} 
Thus an overall renormalization constant $Z_{C}$ is enough  to renormalize this horizon condition. 
  
The effect of this horizon condition to the propagator of the Zwanziger ghost field $\omega_\mu^{AB}(x)$ (see Appendix~B) was calculated based on the localized Gribov-Zwanziger theory in section 10 of  \cite{Zwanziger93}.  
The Zwanziger ghost $\omega$ has the same type of interaction to the gluon and hence it should have the same IR behavior as that of the Faddeev-Popov ghost. 
In this case, the actual calculation is very similar to that for the second horizon term.
It was concluded that the propagator has a $1/(k^2)^2$ singularity at $k=0$. As far as the author understands, however, this conclusion was derived with an implicit assumption ($\lim_{k \rightarrow 0} [k^2 g(k^2)]=0$) similar to $\lim_{k \rightarrow 0}[k^2 v(k^2)]=\lim_{k \rightarrow 0}w(k^2)=w(0)=0$, i.e., absence of a massless pole in a correlation function (10.12) of \cite{Zwanziger93} where $g$ is assumed not to behave as $g(k^2) \sim c/k^2$ for $k^2 \sim 0$ ($c \ne 0$). 
If we remove the implicit assumption, therefore, we reach the same conclusion as that for the second horizon condition given in the previous subsection.
Therefore, we do not pursue this direction. 

\underline{An unconventional method:}
Instead, we work in the original non-local Gribov-Zwanziger formulation. 
We recall the SD equation with the   \textit{horizon condition}:
\begin{equation}
  \quad 
  G^{-1}_{\Lambda}(k^2) =    \frac{G^{-1}_{\Lambda}(0)-2+G_{\Lambda}(0)}{D} -\frac{w_{\Lambda}(0)}{D}  
 -u_{\Lambda}(0)  + u_{\Lambda}(k^2)+w_{\Lambda}(k^2) 
  .
  \label{ghostSD-hc}
\end{equation}
If the horizon condition is incorporated into the SD equation, a partial cancellation at $k=0$ occurs between the horizon condition and the ghost self-energy.  This cancellation always occurs for the multiplicative renormalizable part coming from $\lambda_{\mu\mu}^{AA}(0)$. 
This is not the case for the contribution from the remaining term $\Delta_{\mu\mu}^{AA}(0)$.

In order to avoid the ultraviolet divergence, the UV cutoff $\Lambda$ has been introduced thereby to make the self-energy part $u(k^2)+w(k^2)$ finite. 
Consequently,  $G$ must depend on $\Lambda$, i.e.,  $G(k^2,\Lambda)=G_{\Lambda}(k^2)$. 
In this sense, the above SD equation is an \textit{unrenormalized} version.
The multiplicative renormalization fails for the first horizon term which is non-linear in $G^{-1}$, see (\ref{ghostSD-hc}).

The first horizon condition is not multiplicatively renormalizable. Therefore, the ultraviolet divergence can not be removed within this scheme. 
However, a novel situation occurs by introducing the first horizon condition into the SD equation.
By the resulting combination  $u_{\Lambda}(k^2)-u_{\Lambda}(0)$,   the ultraviolet divergence cancels exactly and  the ultraviolet cutoff $\Lambda$ can be sent to infinity to obtain a finite function of $k^2$. 
Then the above SD equation is regarded as a self-consistent equation to give a finite ghost function $G(k^2)=\lim_{\Lambda \rightarrow \infty}G_{\Lambda}(k^2) < \infty$. 
The term $w_{\Lambda}(k^2)$ is finite from the beginning by some reason, although the $\Lambda$ dependence is apparently assumed. See Appendix~B.
In other words, the SD equation is self-organized (in a non-perturbative way) to give a finite result. 
Therefore there is no need for the specific ultraviolet renormalization in the case of the first horizon term!
Thus, the results obtained in the unrenormalized case hold also after the ultraviolet cutoff $\Lambda$ is send to infinity, as far as the ghost propagator or dressing function is concerned.

For the second horizon condition, a similar  situation does occur.
\begin{equation}
  G^{-1}_{\Lambda}(k^2) =   \frac{1}{D} [1+   u_{\Lambda}(0)] - u_{\Lambda}(0) 
 + u_{\Lambda}(k^2)+w_{\Lambda}(k^2) 
  ,
\end{equation}
which is also rewritten as
\begin{equation}
  G^{-1}_{\Lambda}(k^2) =   \frac{G^{-1}_{\Lambda}(0)}{D}  -   \frac{w_{\Lambda}(0)}{D}    - u_{\Lambda}(0) 
 + u_{\Lambda}(k^2)+w_{\Lambda}(k^2) 
  .
\end{equation}
As already shown, due to the linearity, the UV renormalization of this SD equation can be performed within the multiplicative renormalization framework:
$
 1+ u_{\Lambda}(0)  
= Z_{C}^{-1} [1+ u_R(0) ]
$,
$ 
 w_{\Lambda}(0) = Z_{C}^{-1}w_R(0) 
$
and
$
  G^{-1}_{\Lambda}(0) = Z_{C}^{-1} G^{-1}_R(0)
$.
Moreover, we can use the same argument as the above: The UV divergence cancels for $u(k^2)$ due to the subtraction $- u_{\Lambda}(0)  + u_{\Lambda}(k^2)$, while $w_{\Lambda}(k^2)$ is finite. Therefore, the UV cutoff in $G_{\Lambda}(k^2)$ is sent to infinity without divergences.

Even if the ghost dressing function is ultraviolet finite, we have still finite renormalization coming from the choice of the renormalization point.
Therefore, we study the renormalization point dependence.
From
\begin{align}
  G_R(k^2, \mu^2) 
=&   Z_C^{-1}(\mu^2,\Lambda^2) G(k^2, \Lambda^2)
,
\end{align}
we have
\begin{equation}
 \frac{ G_R(k^2, \mu^2)}{G_R(\mu^2, \mu^2)} 
=     \frac{ G(k^2, \Lambda^2) }{ G(\mu^2, \Lambda^2) }
 .
\end{equation}
In particular, at $k^2=0$ and $\Lambda=\infty$
\begin{equation}
 G_R(0, \mu^2) 
=   G_R(\mu^2, \mu^2)  \frac{ G(0, \infty) }{ G(\mu^2, \infty) }
=   G_R(\mu^2, \mu^2)  \frac{ 1+\sqrt{D}}{ G(\mu^2, \infty) }
 .
\end{equation}
When the decoupling solution with $w(0)=0$ is realized for the first horizon term, 
\begin{equation}
 G_R(0, \mu^2) 
=   G_R(\mu^2, \mu^2) \frac{ G(0)}{ G(\mu^2) }
=  G_R(\mu^2, \mu^2)  \frac{ 1+\sqrt{D}}{ G(\mu^2) }
 .
\end{equation}

For instance, if one chooses the renormalization condition  $G_R (\mu^2, \mu^2) = 1$ 
at $\mu=1.5$GeV for $D=4$, then $G(\mu^2)=1.2$ at $\mu=1.5$GeV for a given boundary condition $G(0)=3$.  In this way, we can reproduce the Orsay data \cite{Boucaudetal08}:
$G_R(0, \mu^2)=3/1.2=2.5$.
The Sao Paulo data \cite{CM08} with our interpretation $G_R(0, \mu^2)=4.2$ and $G_R (\mu^2, \mu^2) = 2$ 
at $\mu=1$GeV is also consistent with this analysis $G(\mu^2)=1.5$ at $\mu=1$. 
Here the value $G(\mu^2)$ is taken from the numerical solution of the Schwinger-Dyson equation or functional renormalization group equation with the boundary condition $G(0)=3$ given in \cite{FMP08} where it is claimed that it is only a matter of infrared boundary conditions $G(0)$ whether scaling or decoupling occurs.

\section{Coulomb gauge}

We consider the Coulomb gauge
\begin{equation}
 \partial_j \mathscr{A}_j(x) = 0 \quad (j=1, \dots, D-1)
 ,
\end{equation}
where $x=({\bf x},x_D)=(x_1,\cdots,x_D)$, ${\bf x}=(x_1,\cdots,x_{D-1})$.
Then the total action reads
\begin{align}
 S_{\rm YM}^{\rm tot}  :=& S_{\rm YM} + S_{\rm GF+FP} ,
 \nonumber\\
 S_{\rm YM} :=&   \int d^Dx \frac14 \mathscr{F}_{\mu\nu}  \cdot \mathscr{F}_{\mu\nu} ,
\quad
 S_{\rm GF+FP} := \int d^Dx  \left\{ \mathscr{B} \cdot \partial_j \mathscr{A}_j 
+i \bar {\mathscr{C}} \cdot \partial_j D_j \mathscr{C} \right\} 
 ,
\end{align}
We can adopt the horizon term which is instantaneous in $x_D=y_D :=t$: \cite{Zwanziger06}
\begin{equation}
 h[\mathscr{A}(x)] 
=  \int d^{D-1}{\bf y} gf^{ABC} \mathscr{A}_j^{B}(x) (K^{-1})^{CE}(x,y) gf^{AFE} \mathscr{A}_j^{F}(y) |_{x_D=y_D}  , 
\label{hC1}
\end{equation}
or
\begin{equation}
  h[\mathscr{A}(x)] =  \int d^{D-1}{\bf y} D_j[\mathscr{A}]^{AC}(x) (K^{-1})^{CE}(x,y) D_j[\mathscr{A}]^{AE}(y)|_{x_D=y_D} ,
\label{hC2}
\end{equation}
where $K$ is the Faddeev-Popov operator in the Coulomb gauge $K[\mathscr{A}]:=-\partial_j D_j=-\partial_j (\partial_j+g \mathscr{A}_j \times)$.
The horizon action $S_h$ is non-local in space, but local in time. 
The horizon condition is given by 
\begin{equation}
 \langle h(x) \rangle_{\rm GZ} = (N^2-1)(D-1) .
\end{equation}
Here $\langle h(x) \rangle_{\rm GZ}$ is  independent of $t$ due to translational invariance. 
The horizon action $S_h$ is non-local in space, but local in time. For instance, 
\begin{align}
S_h =& \gamma \int d^{D}x h[\mathscr{A}(x)] 
=  \gamma  \int dt \int d^{D-1}{\bf x} h[\mathscr{A}(x)] 
\nonumber\\
=&   \gamma  \int dt \int d^{D-1}{\bf x} \int d^{D-1}{\bf y} gf^{ABC} \mathscr{A}_j^{B}(x) (K^{-1})^{CE}(x,y) gf^{AFE} \mathscr{A}_j^{F}(y) |_{x_D=y_D}  . 
\end{align}
The ghost propagator (i.e., the complete propagator of the Faddeev-Popov ghost) $
G^{AB}(x-y) = \langle \mathscr{C}^A(x) \bar{\mathscr{C}}^B(y)  \rangle 
$
is instantaneous,
$
 G^{AB}(x) = G^{AB}({\bf x}) \delta(t)
$.
In momentum space, it is independent of $k_D$, i.e.,
\begin{equation}
   G^{AB}({\bf k}, k_D) = G^{AB}({\bf k}) .
\end{equation}
The free ghost propagator is independent of $k_D$
\begin{equation}
    G^{AB}_{0}({\bf k}) = \delta^{AB} \frac{1}{{\bf k}^2} .
\end{equation}
We can introduce the ghost dressing function (or form factor) $d({\bf k}^2)$ which is dimensionless by
\begin{equation}
   G^{AB}({\bf k}, k_D) = G^{AB}({\bf k}) = -\delta^{AB} \frac{d({\bf k}^2)}{{\bf k}^2} .
\end{equation}

In the similar way, we can show that the identity holds:
\begin{equation}
    d({\bf k}^2) = [1+u({\bf k}^2)+w({\bf k}^2)]^{-1} .
\end{equation}
It is also shown that the first horizon term (\ref{hC1}) yields  
\begin{align}
 \langle  h(0) \rangle
  =& - \lim_{{\bf k} \to {\bf 0}}  \langle   (g \mathscr{A}_j \times \mathscr{C})^A (g \mathscr{A}_j \times  \bar{\mathscr{C}})^A \rangle_{{\bf k}}
\nonumber\\
=& 
 - (N^2-1)  \left\{ (D-1)u({\bf 0}) +w({\bf 0})- d({\bf 0}) [u({\bf 0})+w({\bf 0})]^2  \right\}  
 ,
\end{align} 
while the second horizon function (\ref{hC2}) leads to
\begin{align}
 \langle  h(0) \rangle
=  -\lim_{{\bf k}  \rightarrow {\bf 0}} \langle   (D_j  \mathscr{C})^A (D_j   \bar{\mathscr{C}})^A \rangle_{{\bf k}} 
= -(N^2-1) 
  \left\{ (D-2)u({\bf 0}) -1 \right\}  
 ,
\end{align} 
where $u({\bf k}^2)$ and $w({\bf k}^2)$ are defined by 
\begin{equation}
 \lambda_{ij}^{AB}({\bf k}) := \langle   (g \mathscr{A}_i \times \mathscr{C})^A (g \mathscr{A}_j \times  \bar{\mathscr{C}})^B \rangle_{{\bf k}}^{m1PI}
=  \left[  \delta_{ij} u({\bf k}^2) + \frac{{\bf k}_i {\bf k}_j}{{\bf k}^2} w({\bf k}^2) \right] \delta^{AB} 
 .
\end{equation}
Thus, we can obtain the similar results in the Coulomb gauge to those in the Landau gauge by replacing $D$ in the Landau gauge with $D-1$ in the Coulomb gauge. 
We find one-parameter family of solutions including both the scaling and decoupling solutions.
See e.g. \cite{LM04} for the ghost dressing function obtained by the numerical simulations on lattice.

\section{Conclusion and discussion}

In this paper we have discussed how the existence of the Gribov horizon affects the deep infrared behavior of the ghost propagator in the Landau and Coulomb gauge $G=SU(N)$ Yang-Mills theory, using the Gribov-Zwanziger framework with the horizon condition $\langle h(x) \rangle = ({\rm dim}G)D$. 
 Moreover, we have shown how to incorporate the horizon condition into the Schwinger-Dyson equation for the ghost propagator to discriminate between scaling and decoupling. 
 We have examined two horizon conditions derived from two types of horizon terms, both of which were proposed by Zwanziger \cite{Zwanziger89,Zwanziger93}. 
 We have shown that one parameter family of solutions parameterized by $w(0)$ exists in both cases, although some results crucially depend on the choice  of the horizon term adopted. 
 The value $w(0)$ has been assumed implicitly to be zero $w(0)=0$ in previous studies.

For the first horizon term \cite{Zwanziger89}, 
\begin{equation}
 h(x) 
=  \int d^Dy gf^{ABC} \mathscr{A}_\mu^{B}(x) (K^{-1})^{CE}(x,y) gf^{AFE} \mathscr{A}_\mu^{F}(y)  , 
\end{equation}
the GZ theory is not multiplicatively renormalizable. 
However, the SD equation for the ghost propagator and the ghost dressing function can be UV finite, once the horizon condition is incorporated into the SD equation as proposed in this paper.  
The decoupling solution, i.e., finite ghost dressing function $G(k^2)$ even in the limit $k \rightarrow 0$: $G(0) < \infty$ is allowed to exist, unless $w(0)=-\infty$.
The Kugo-Ojima criterion $u(0) = -1$ is not necessarily satisfied except for a special choice of $w(0)=1/2$ for any $D$, leading to $G(0)=2$.
For $w(0)=0$, the boundary values are $G(0)=1+\sqrt{D}$ and  $u(0)=(-D + \sqrt{D})/(D-1)$  up to renormalization point dependence. 
A possible renormalization scheme and the renormalization point dependence of the decoupling solution has been  discussed.
This should be compared with the paper \cite{ABP09}.       

For the second horizon term \cite{Zwanziger93}
\begin{equation}
  h(x) =  \int d^Dy D_\mu[\mathscr{A}]^{AC}(x) (K^{-1})^{CE}(x,y) D_\mu[\mathscr{A}]^{AE}(y) ,
\end{equation}
the GZ theory is multiplicatively renormalizable. 
The Kugo-Ojima criterion $u(0) = -1$ is satisfied both in the unrenormalized and the renormalized cases \cite{NF00,Dudal09a}. 
For $w_R(0)=0$,  the scaling solution, i.e., infinite ghost dressing function in the  limit $k \rightarrow 0$: $G_R(0) = w_R(0)^{-1} = \infty$, even after the renormalization.
For  $w_R(0) \ne 0$, the decoupling solution is obtained against the claim in the previous literatures. 
 
Thus the investigation of  $w_R(0)$ is crucial to see which solution is realized.  
An interesting step towards this direction was done in \cite{ABP09}. However, according to our analysis, their result, namely, the  decoupling solution with $w_R(0)=0$ is not compatible with the multiplicative renormalization scheme they used: 
$
 1+ u(0)  
= Z_{C}^{-1} [1+ u_R(0) ]
$,
$ 
 w(0) = Z_{C}^{-1}w_R(0) 
$
and
$
  G^{-1}(0) = Z_{C}^{-1} G^{-1}_R(0)
$.
This issue should be reexamined and confirmed by further investigations. 

If we require that two horizon conditions give the same result, then the relation
$
 u(0)=-1/2-w(0)
$
must be satisfied for any $D$.
This implies that for any $D$
\begin{equation}
  G(0)  =2, \quad   u(0) =  -1 , \quad w(0)= \frac12  
  .
\end{equation}
Then, in the unrenormalized case, the Kugo-Ojima criterion is compatible with the decoupling solution for both choice of horizon term, against the conventional wisdom where the scaling solution is believed to be consistent with the Kugo-Ojima criterion. 
However, $G(0)$, $u(0)$ and $w(0)$ are not renormalization group invariants.  There is no guarantee that this relation is preserved. In fact, the Kugo-Ojima criterion is preserved only for the second horizon term. 
In order to judge which horizon function is realized, one need to know the result of numerical simulations on finer lattice, i.e.,  with smaller lattice spacing corresponding to larger ultraviolet cutoff, in addition to   larger size lattices. 
But this might be impossible, because the horizon term alone does not uniquely fix the gauge, since there are still Gribov copies in the first Gribov region.

Thus we can conclude that $w_R(0)$ plays the role of an additional non-perturbative gauge parameter which uniquely specifies the solution from a one-parameter family of solutions including the scaling and decoupling.  
In view of this, the value of $w_R(0)$  itself   has no physical meaning. 
In fact, it has been shown \cite{BGP07} that all solutions (decoupling as well as scaling) lead to quark confinement by proving the vanishing of  the Polyakov loop as a gauge-invariant order parameter of quark confinement.
In this sense, discriminating between decoupling and  scaling may not be so important from the physical point of view and main contribution to phenomenological studies comes from the high-momentum region and the intermediate region around 1GeV which is stable irrespective of adopting the scaling or decoupling solution. 
It is shown that both scaling and decoupling solutions do not contradict the general principles of quantum gauge field theories \cite{Kondo04}.

How the existence of the horizon is relevant for color confinement. 
In the Gribov-Zwanziger theory (restricted to the 1st Gribov region), the BRST symmetry is broken by the existence of the horizon. 
$
\mbox{\boldmath $\delta$} S_{\rm GZ}
=\mbox{\boldmath $\delta$} \tilde S_\gamma \ne 0
$.
Nevertheless, there exists  a ``BRST'' like symmetry (without nilpotency \cite{Sorella09}  or with nilpotency \cite{Kondo09b} which leaves the Gribov-Zwanziger action invariant.  
Then we could apply the Kugo-Ojima idea to the Gribov-Zwanziger theory, which opens the path to searching for the modified color confinement criterion {\it a la} Kugo and Ojima.  
In view of this, defining a non-perturbative BRST transformation will be another interesting possibility to be investigated   \cite{Smekal08}.

\vskip 1cm
[Note added]
\vskip 0.3cm
In preparing this paper, it is pointed out in \cite{Boucaudetal09}   that by writing $u(0)$ and $w(0)$ as function of $G(0)$,  $u_{\Lambda}(0) \rightarrow +\infty$ and $w_{\Lambda}(0) \rightarrow -\infty$  such that $u_{\Lambda}(0)+w_{\Lambda}(0) \rightarrow -1$, provided that  $G_{\Lambda}(0) \rightarrow \infty$ as $\Lambda \rightarrow \infty$, see Appendix~C.
This analysis uses the first horizon condition and the relation $G(0)^{-1}=1+u(0)+w(0)$. However, the statement $G_{\Lambda}(0) \rightarrow \infty$ as $\Lambda \rightarrow \infty$ is a result of perturbation theory.  This analysis does not use the full information coming from the relation $G(k^2)^{-1}=1+u(k^2)+w(k^2)$  (the Schwinger-Dyson equation for the ghost dressing function)  for the whole momentum region.  As we have shown in the text, $G_{\Lambda}(0)$ remains finite even after the limit $\Lambda \rightarrow \infty$ in a non-perturbative way, once the full information  is used.  

\section*{Acknowledgments}
The author would like to thank Christian Fischer, Lorenz von Smekal and Jan Pawlowski for valuable discussions and warm hospitality at Institut f\"ur kernphysik, Technische Universit\"at Darmstadt, Eiji Nakano, Bengt Friman and  Krzysztof Redlich for helpful discussions and kind hospitality at GSI (Gesellschaft fur Schwerionenforschung),
and Olivier Pene,  Jose Rodriguez-Quintero,  David Dudal, Andrea Quadri and Hugo Reinhardt for useful discussions and Danielle Binosi, Joannis Papavassiliou, John Cornwall and  Arlene Aguilar for hospitality at ECT*, Trento. 
He is grateful to High Energy Physics Theory Group and Theoretical Hadron Physics Group in the University of Tokyo, especially, Prof. Tetsuo Hatsuda for kind hospitality extended to him on sabbatical leave.
This work is financially supported by  Grant-in-Aid for Scientific Research (C) 21540256 from Japan Society for the Promotion of Science
(JSPS).

\appendix
\section{The function $w$ and its ultraviolet behavior}

Two functions $u$ and $w$ are defined by
\begin{equation}
\lambda_{\mu\nu}^{AB}(k)
:= \langle   (g \mathscr{A}_\mu \times \mathscr{C})^A (g \mathscr{A}_\nu \times  \bar{\mathscr{C}})^B \rangle_k^{\rm m1PI}
=  \left[  \delta_{\mu\nu} u(k^2) + \frac{k_\mu k_\nu}{k^2} w(k^2) \right] \delta^{AB} 
 , 
\end{equation}
which implies
\begin{subequations}
\begin{align}
 u(k^2) =& \frac{1}{(D-1)(N^2-1)} \left[  \delta^{\mu\nu} - \frac{k^\mu k^\nu}{k^2}  \right] \lambda_{\mu\nu}^{AA}(k)
  ,
  \label{eq-u}
  \\
 w(k^2) =& \frac{-1}{(D-1)(N^2-1)} \left[  \delta^{\mu\nu} - D \frac{k^\mu k^\nu}{k^2}    \right] \lambda_{\mu\nu}^{AA}(k)
  .
  \label{eq-w}
\end{align}
\end{subequations}
The argument given in \cite{GHQ04} is based on the power counting. 
$
 {\rm dim.}[\mathscr{A}]=(D-2)/2={\rm dim.}[\mathscr{C}]={\rm dim.}[\bar{\mathscr{C}}]
$ means
$
{\rm dim.}[\mathscr{A}\mathscr{C}\mathscr{A}\bar{\mathscr{C}}]=2D-4
$.
${\rm dim.}[g]=(4-D)/2$.
Therefore, 
$
{\rm dim.}[ \left< \mathscr{A}\mathscr{C}\mathscr{A}\bar{\mathscr{C}} \right>_{k} ]= 2D-4-D=D-4
$.
Thus, for $D=4$,
$\lambda_{\mu\nu}^{AB}(k)$ has at most logarithmic  divergent.
Only $u$ has divergence, while $w$ is ultraviolet finite. 
The 1PI part has no massless pole.
The ultraviolet divergence appears in $u$ and not in $w$. So the ultraviolet divergence is proportional to $g_{\mu\nu}$.
In fact, the Brown-Pennington projector $\left[ g^{\mu\nu} - D \frac{k^\mu k^\nu}{k^2}  \right]$ eliminates the term proportional to $g_{\mu\nu}$.

\section{A localized Gribov-Zwanziger theory}

The Gribov-Zwanziger theory can be rewritten into the local form \cite{Zwanziger93} by introducing additional fields called the Zwanziger ghosts $\xi,\bar{\xi},\omega,\bar{\omega}$:
\begin{equation}
   e^{-\gamma \int d^D x h(x) } 
=
\int [d\xi] [d\bar{\xi}]  [d\omega] [d\bar{\omega}]  
 \exp \left\{   - \tilde{S}_\gamma[\mathscr{A},\xi,\bar{\xi},\omega,\bar{\omega}]  \right\} 
 ,
\end{equation}
where
\begin{align}
    \tilde{S}_\gamma  
=: \int d^Dx & [
  \bar{\xi}_\mu^{CA} K^{AB} \xi_\mu^{CB} 
-  \bar{\omega}_\mu^{CA} K^{AB} \omega_\mu^{CB}
\nonumber\\&
+ i \gamma^{1/2} gf^{ABC} \mathscr{A}_\mu^B  \xi_\mu^{AC} + i \gamma^{1/2} gf^{ABC} \mathscr{A}_\mu^B \bar{\xi}_\mu^{AC} 
 ] 
  .  
  \label{aux-action}
\end{align}
The localized action $S_{\rm GZ}$ for the Gribov-Zwanziger theory is obtained
\begin{align}
S_{\rm GZ}  =&  S_{\rm YM}^{\rm tot}[\mathscr{A},\mathscr{C},\bar{\mathscr{C}},\mathscr{B}] + \tilde{S}_\gamma [\mathscr{A},\xi,\bar{\xi},\omega,\bar{\omega}]  
  \nonumber\\
=& S_{\rm YM}[\mathscr{A}] + S_{\rm GF+FP}[\mathscr{A},\mathscr{C},\bar{\mathscr{C}},\mathscr{B}] 
+ \tilde{S}_\gamma [\mathscr{A},\xi,\bar{\xi},\omega,\bar{\omega}]
 ,
\end{align}
where
\begin{align}
 \mathscr{L}_{\rm GF+FP}  :=& \int d^Dx  \left\{ \mathscr{B} \cdot \partial_\mu \mathscr{A}_\mu 
+i \bar {\mathscr{C}} \cdot \partial_\mu D_\mu \mathscr{C} \right\} 
  .
\end{align}

The localized GZ theory is known to be multiplicatively renormalizable to all orders where the multiplicative renormalization factors are introduced as 
\begin{align}
 \mathscr{A}_\mu =& Z_{A}^{1/2} \mathscr{A}_\mu^{R} , \quad
 \mathscr{B} = Z_{B}^{1/2} \mathscr{B}^{R} , \quad Z_{B} = Z_{A}^{-1} 
 ,
 \nonumber\\
 \mathscr{C} =& Z_{C}^{1/2} \mathscr{C}^{R} , \quad
 \bar{\mathscr{C}} = Z_{C}^{1/2} \bar{\mathscr{C}}^{R}  ,
 \nonumber\\
 g =& Z_g g_R , \quad Z_g = \tilde{Z}_1 Z_{A}^{-1/2} Z_{C}^{-1} 
,
\label{ren-const}
\end{align}
and
\begin{align}
 \xi_\mu =& Z_{\xi}^{1/2} \xi_\mu^{R} , \quad
 \bar\xi_\mu = Z_{\bar\xi}^{1/2} \bar\xi_\mu^{R} , \quad
 Z_{\xi} = Z_{\bar\xi} = Z_{C}
  ,
 \nonumber\\
 \omega_\mu =& Z_{\omega}^{1/2} \omega_\mu^{R} , \quad
 \bar\omega_\mu = Z_{\bar\omega}^{1/2} \bar\omega_\mu^{R} , \quad\ 
 Z_{\omega} = Z_{\bar\omega} = Z_{C}
  ,
 \nonumber\\
 \gamma =& Z_\gamma \gamma_R , \quad Z_\gamma = Z_{A}^{-1} Z_{C}^{-1}
 .
\end{align}
In the localized GZ action,  the covariant derivative form can be used
\begin{equation}
    \tilde{S}_\gamma  
= \int d^Dx   \{
  \bar{\xi}_\mu^{CA} K^{AB} \xi_\mu^{CB} 
-  \bar{\omega}_\mu^{CA} K^{AB} \omega_\mu^{CB}
+ i \gamma^{1/2} D_\mu[\mathscr{A}]^{AC}  \xi_\mu^{AC} + i \gamma^{1/2} D_\mu[\mathscr{A}]^{AC} \bar{\xi}_\mu^{AC} 
 \} 
  .  
\end{equation}
This means another form of the non-local horizon term 
\begin{equation}
 h(x) 
=  \int d^Dy D[\mathscr{A}]_\mu^{AC}(x) (K^{-1})^{CE}(x,y) D[\mathscr{A}]_\mu^{AE}(y)
 . 
\end{equation}

\section{$u(0)$ and $w(0)$ as functions of $G(0)$}

\noindent
The analysis done in \cite{Boucaudetal09} is peformed to two horizon terms to compare their implications. 

For the first horizon term
$
 \frac{\langle h(0) \rangle}{D(N^2-1)}=D^{-1}\{-(D-1)u(0)-G(0)[u(0)+w(0)]\}
$,
\begin{align}
u_{\Lambda}(0) &= \frac{1}{D-1} \left\{ G_{\Lambda}(0)-1   -  D \left[\frac{\langle h(0) \rangle}{D(N^2-1)} \right] \right\} ,
\nonumber\\
w_{\Lambda}(0) 
 &=   \frac{1}{D-1} \left\{ - G_{\Lambda}(0)+2-D + D \left[\frac{\langle h(0) \rangle}{D(N^2-1)} \right] \right\} + \frac{1}{G_{\Lambda}(0)} .
\end{align}
Using the horizon condition,
\begin{align}
u_{\Lambda}(0) &= \frac{1}{D-1} \left\{ G_{\Lambda}(0)-1   -  D  \right\} ,
\nonumber\\
w_{\Lambda}(0) 
 &=   \frac{1}{D-1} \left\{ -G_{\Lambda}(0)+2  \right\} + \frac{1}{G_{\Lambda}(0)} .
\end{align}
If $G_{\Lambda}(0) \rightarrow \infty$ as $\Lambda \rightarrow \infty$, then
$u_{\Lambda}(0) \rightarrow +\infty$ and $w_{\Lambda}(0) \rightarrow -\infty$  such that $u_{\Lambda}(0)+w_{\Lambda}(0) \rightarrow -1$.

\begin{figure}
\begin{center}
\includegraphics[width=8cm]{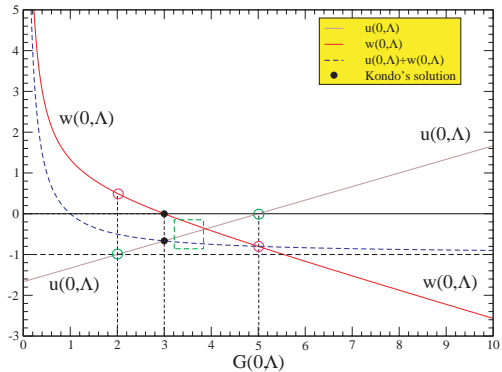}
\end{center}
\caption{{\small 
This graph is extracted from \cite{Boucaudetal09}.
 The solutions for $u(0,\Lambda)$ and $w(0,\Lambda)$ plotted as a function of $G(0,\Lambda)$. 
The particular solution, $G(0)=3$, $u(0)=-2/3$, proposed by Kondo (black circles), obtained  by imposing $w(0,\Lambda)=0$, corresponds to the intersection of $u+w$ and $u$. 
 The current lattice solutions for the bare ghost dressing functions at vanishing momentum lie inside the green  dotted square. The Kugo-Ojima parameter is in the region $-1<u(0)<0$, when $2<G(0)<5$.
}}
\label{kondo-plot}
\end{figure}

For the second horizon term
$
 \frac{\langle h(0) \rangle}{D(N^2-1)}=D^{-1}\{-(D-1)u(0)+1\}
$
\begin{align}
u_{\Lambda}(0) &= \frac{1}{D-1} \left\{ 1   -  D \left[\frac{\langle h(0) \rangle}{D(N^2-1)} \right] \right\} ,
\nonumber\\
w_{\Lambda}(0) 
 &= - \frac{1}{D-1} \left\{ D - D \left[\frac{\langle h(0) \rangle}{D(N^2-1)} \right] \right\} + \frac{1}{G_{\Lambda}(0)} .
\end{align}
Using the horizon condition,
\begin{align}
u_{\Lambda}(0) &= -1 \ (\Lambda-\text{indep.}) ,
\nonumber\\
w_{\Lambda}(0) 
 &=  \frac{1}{G_{\Lambda}(0)} .
\end{align}
If $G_{\Lambda}(0) \rightarrow \infty$ as $\Lambda \rightarrow \infty$, then
$w_{\Lambda}(0) \rightarrow 0$  such that $u_{\Lambda}(0)+w_{\Lambda}(0) \rightarrow -1$.
It should be remarked that $u_{\Lambda}(0) =-1$ independently of $\Lambda$. This result seems to be inconsistent with the lattice result \cite{FN07,Sternbeck06}.

\noindent
Remark: 
In the first horizon term, if the contribution from $\Delta_{\mu\mu}$ is neglected \cite{Zwanziger09}, then
$
 \frac{\langle h(0) \rangle}{D(N^2-1)}=D^{-1}\{-Du(0)-w(0) \}
$,
\begin{align}
u_{\Lambda}(0) &= \frac{1}{D-1} \left\{ 1- \frac{1}{G_{\Lambda}(0)}   -  D \left[\frac{\langle h(0) \rangle}{D(N^2-1)} \right] \right\} ,
\nonumber\\
w_{\Lambda}(0) 
 &=   \frac{1}{D-1} \left\{ -D+D\frac{1}{G_{\Lambda}(0)} + D \left[\frac{\langle h(0) \rangle}{D(N^2-1)} \right] \right\} .
\end{align}
Using the horizon condition,
\begin{align}
u_{\Lambda}(0) &= \frac{1}{D-1} \left\{ 1- \frac{1}{G_{\Lambda}(0)}   -  D  \right\} <-1 ,
\nonumber\\
w_{\Lambda}(0) 
 &=  \frac{D}{D-1}   \frac{1}{G_{\Lambda}(0)} .
\end{align}
If $G_{\Lambda}(0) \rightarrow \infty$ as $\Lambda \rightarrow \infty$, then
$u_{\Lambda}(0) \rightarrow -1$ and $w_{\Lambda}(0) \rightarrow 0$  such that $u_{\Lambda}(0)+w_{\Lambda}(0) \rightarrow -1$.  However, $u_{\Lambda}(0) <-1$ which contradicts with the lattice result.



\baselineskip 12pt

\begin{thebibliography}{999}
\bibitem{YM54}
  C.N. Yang and R.L. Mills,
Phys. Rev. {\bf 96}, 191--195  (1954).


\bibitem{Gribov78}
 V.N. Gribov,
Nucl. Phys. B{\bf 139}, 1--19 (1978).
 
 
\bibitem{AS01}
R. Alkofer and L. von Smekal,
hep-ph/0007355,
Phys.Rept.{\bf 353}, 281 (2001)]
\\
L. von Smekal, R. Alkofer, A. Hauck, 
e-Print: hep-ph/9705242,
Phys.Rev.Lett.{\bf 79}, 3591-3594 (1997).


\bibitem{KO79}
 T. Kugo and I. Ojima,
Suppl. Prog. Theor. Phys. {\bf 66}, 1--130 (1979).

 
\bibitem{Kugo95}
 T. Kugo,
hep-th/9511033.


\bibitem{Boucaudetal08}
 Ph. Boucaud, J.P. Leroy, A. Le Yaouanc, J. Micheli, O. Pene and J. Rodriguez-Quintero, 
hep-ph/0803.2161,
JHEP {\bf 06}, 099 (2008).
\\
 Ph. Boucaud, J.P. Leroy, A. Le Yaouanc, J. Micheli, O. Pene and J. Rodriguez-Quintero, 
arXiv:0801.2721[hep-ph],
JHEP {\bf 06}, 012 (2008).


\bibitem{LS02}
C. Lerche and  L. von Smekal, 
e-Print: hep-ph/0202194,
Phys.Rev.D{\bf 65}, 125006 (2002). 


\bibitem{SIMPS06}
A. Sternbeck, E.-M. Ilgenfritz, M. M\"uller-Preussker, A. Schiller and I.L. Bogolubsky,
hep-lat/0610053, 
Talk given at 24th International Symposium on Lattice Field Theory (Lattice 2006), Tucson, Arizona, 23-28 Jul 2006. 
Published in PoS LAT2006:076,2006. PoS LAT2006:076,2006 (Lattice 2006).


\bibitem{BMMP08}
V.G. Bornyakov, V.K. Mitrjushkin and M. M\"uller-Preussker,
arXiv:0812.2761[hep-lat].


\bibitem{BIMPS09}
I.L. Bogolubsky, E.-M. Ilgenfritz and M. M\"uller-Preussker and A. Sternbeck,
arXiv:0901.0736[hep-lat],
Phys. Lett. B {\bf 676}, 69--73 (2009).


\bibitem{CM07}
A. Cucchieri, T. Mendes, O. Oliveira and P.J. Silva, 
e-Print: arXiv:0705.3367 [hep-lat],  
Phys. Rev. D{\bf 76}, 114507 (2007). 
 
 
\bibitem{CM08}
 A. Cucchieri and T. Mendes,
arXiv:0804.2371[hep-lat],
Phys. Rev. D{\bf 78}, 094503 (2008),
\\
 A. Cucchieri and T. Mendes,
arXiv:0812.3261[hep-lat].
\\
 A. Cucchieri and T. Mendes,
arXiv:0904.4033[hep-lat].


\bibitem{OS08}
O. Oliveira and P.J. Silva,
arXiv:0809.0258[hep-lat].


\bibitem{SS08}
 A. Sternbeck and L. von Smekal,
arXiv:0811.4300[hep-lat].
\\
A. Sternbeck, L. von Smekal, D.B. Leinweber, A.G. Williams,
 arXiv:0710.1982[hep-lat],
Pos LAT2007, 340 (2007). 


\bibitem{FMP08}
C.S.~Fischer, A.~Maas and J.M.~Pawlowski,
arXiv:0810.1987 [hep-ph],
Annals of Physics,  in press.


\bibitem{FMP07}
C.S. Fischer, J.M. Pawlowski, 
e-Print: hep-th/0609009, 
Phys.Rev. D{\bf 75}, 025012 (2007). 
\\
C.S. Fischer, J.M. Pawlowski, 
e-Print: arXiv:0903.2193 [hep-th], 
Phys.Rev. D{\bf 80}, 025023 (2009). 


\bibitem{Maas09}
A. Maas, 
e-Print: arXiv:0907.5185 [hep-lat]. 


\bibitem{ABP08}
 A.C. Aguilar, D. Binosi and J. Papavassiliou,
arXiv:0802.1870 [hep-ph],
Phys. Rev. D{\bf 78}, 025010 (2008).
\\
 A.C. Aguilar, D. Binosi and J. Papavassiliou,
PoS LC2008:050,2008
arXiv:0810.2333 [hep-ph].


\bibitem{Dudaletal08}
 D. Dudal, J.A. Gracey, S.P. Sorella, N. Vandersickel and H. Verschelde,
arXiv:0806.4348[hep-th],
Phys. Rev. D{\bf 78}, 065047 (2008).
\\
 D. Dudal, S.P. Sorella, N. Vandersickel and H. Verschelde,
arXiv:0711.4496[hep-th],
Phys. Rev. D{\bf 77}, 071501(R) (2008).


\bibitem{Zwanziger89}
 D. Zwanziger,
Nucl. Phys. B{\bf 323}, 513--544 (1989).
 

\bibitem{Zwanziger92}
 D. Zwanziger,
Nucl. Phys. B{\bf 378}, 525--590 (1992).


\bibitem{Zwanziger93}
 D. Zwanziger,
Nucl. Phys. B{\bf 399}, 477--513 (1993).
 

\bibitem{Zwanziger94}
D. Zwanziger,
Nucl.Phys.B{\bf 412}, 657-730 (1994). 


\bibitem{Kondo09a}
 K.-I. Kondo,
arXiv:0904.4897 [hep-th],
Phys.Lett.B. {\bf 678}, 322-330 (2009).


\bibitem{Kondo09c}
 K.-I. Kondo,
arXiv:0907.3249 [hep-th],
Prog. Theor. Phys. to be published.


\bibitem{GHQ04}
P.A. Grassi, T. Hurth, A. Quadri, 
e-Print: hep-th/0405104, 
Phys.Rev. D{\bf 70}, 105014 (2004). 


\bibitem{Zwanziger06}
D. Zwanziger,
Phys. Rev. D{\bf 76}, 125014 (2007).
\\
D. Zwanziger,
Braz. J. Phys. {\bf 37}, 127--143 (2007).


\bibitem{LM04}
K. Langfeld and L. Moyaerts,
hep-lat/0406024, 
Phys. Rev. D{\bf 70}, 074507 (2004).


\bibitem{ABP09}
A.C. Aguilar, D. Binosi and J. Papavassiliou,
arXiv:0907.0153 [hep-ph]. 
\\
A.C. Aguilar, D. Binosi, J. Papavassiliou and J. Rodriguez-Quintero, 
arXiv:0906.2633 [hep-ph] 


\bibitem{NF00}
H. Nakajima and S. Furui, 
hep-lat/0006002, 
Talk given at International Symposium on Quantum Chromodynamics (QCD) and Color Confinement (Confinement 2000).  Published in *Osaka 2000, Quantum chromodynamics and color confinement* 60-69 (World Scientific, Singapore). 


\bibitem{FN07}
 S. Furui and H. Nakajima,
hep-lat/0609024,
Brazilian Journal of Physics {\bf 37}, 186--192 (2007).
\\
 S. Furui and H. Nakajima,
hep-lat/0503029,
Few-Body Systems {\bf 40}, 101--128 (2006). 
\\
 S. Furui and H. Nakajima,
hep-lat/0305010,
Phys. Rev. D{\bf 69}, 074505 (2004). 


\bibitem{Sternbeck06}
A. Sternbeck,
section 5.2 in hep-lat/0609016. 


\bibitem{Maas08}
 A. Maas,
 arXiv:0808.3047[hep-lat],
Phys. Rev. D{\bf 79}, 014505 (2009).


\bibitem{CMM04}
A. Cucchieri, T. Mendes, A. Mihara,
e-Print: hep-lat/0408034, 
JHEP {\bf 04}, 012 (2004). 


\bibitem{IMPSS06}
E.-M. Ilgenfritz, M. Muller-Preussker, A. Sternbeck and A. Schiller, 
e-Print: hep-lat/0601027 


\bibitem{Dudal09a}
 D. Dudal, S.P. Sorella, N. Vandersickel and H. Verschelde,
arXiv:0904.0641[hep-th],
Phys.Rev.D{\bf 79}, 121701 (2009). 


\bibitem{BGP07}
J. Braun, H. Gies and J.M. Pawlowski, 
e-Print: arXiv:0708.2413 [hep-th]].


\bibitem{Kondo04}
K.-I. Kondo, 
e-Print: hep-th/0303251.
\\
K.-I. Kondo, 
e-Print: hep-lat/0309142, 
Nucl. Phys. Proc. Suppl. {\bf 129}, 715-717 (2004).


\bibitem{Sorella09}
S.P. Sorella,
arXiv:0905.1010[hep-th],
%
Phys.Rev.D{\bf 80}, 025013 (2009).


\bibitem{Kondo09b}
 K.-I. Kondo,
arXiv:0905.1899[hep-th].


\bibitem{Smekal08}
L. von Smekal,  
e-Print: arXiv:0812.0654 [hep-th]. 
\\
L. von Smekal, A. Jorkowski, D. Mehta, A. Sternbeck, 
e-Print: arXiv:0812.2992 [hep-th]. 
\\
H. Neuberger, 
Phys.Lett. B{\bf 183}, 337 (1987).

\bibitem{Zwanziger09}
 D. Zwanziger,
arXiv:0904.2380[hep-th].


\bibitem{Boucaudetal09}
Ph. Boucaud, J.P. Leroy, A.Le Yaouanc, J. Micheli, O. Pene and J. Rodriguez-Quintero, 
e-Print: arXiv:0909.2615 [hep-ph].



\end{thebibliography}
\end{document}